\documentclass[fleqn,usenatbib]{mnras}

\usepackage[T1]{fontenc}

\DeclareRobustCommand{\VAN}[3]{#2}
\let\VANthebibliography\thebibliography
\def\thebibliography{\DeclareRobustCommand{\VAN}[3]{##3}\VANthebibliography}

\usepackage{booktabs}
\usepackage{amsmath} 
\usepackage{array}   
\usepackage{multirow} 
\usepackage{newtxtext, newtxmath}
\usepackage{kotex} 
\usepackage{makecell} 
\usepackage{hyperref} 
\usepackage{needspace}
\usepackage{graphicx}	
\usepackage{longtable}
\usepackage{comment}
\usepackage{bm}

\defcitealias{2011ApJ...740...92G}{G11}
\defcitealias{2019ApJ...874...32R}{R19}
\defcitealias{2021MNRAS.503L..33Z}{Z21}

\title[Age-bias in SN Cosmology. II.]{Strong Progenitor Age-bias in Supernova Cosmology. II.\\Alignment with DESI BAO and Signs of a Non-Accelerating Universe}

\author[J. Son et al.]{
Junhyuk Son,\thanks{E-mail: sonjunhyuk@yonsei.ac.kr (JHS)}
Young-Wook Lee,\thanks{E-mail: ywlee2@yonsei.ac.kr (YWL)}
Chul Chung,\thanks{E-mail: chulchung@yonsei.ac.kr (CC)}
Seunghyun Park
and Hyejeon Cho
\\
Department of Astronomy \& Center for Galaxy Evolution Research, Yonsei University, Seoul 03722, Republic of Korea\\
}

\date{Accepted XXX. Received YYY; in original form ZZZ}

\pubyear{\the\year{}}

\begin{document}
\label{firstpage}
\pagerange{\pageref{firstpage}--\pageref{lastpage}}
\maketitle

\begin{abstract}
Supernova (SN) cosmology is based on the key assumption that the luminosity standardization process of Type Ia SNe remains invariant with progenitor age. However, direct and extensive age measurements of SN host galaxies reveal a significant ($5.5 \sigma$) correlation between standardized SN magnitude and progenitor age, which is expected to introduce a serious systematic bias with redshift in SN cosmology. This systematic bias is largely uncorrected by the commonly used mass-step correction, as progenitor age and host galaxy mass evolve very differently with redshift. After correcting for this age-bias as a function of redshift, the SN dataset aligns more closely with the $w_0w_a$CDM model recently suggested by the DESI BAO project from a combined analysis using only BAO and CMB data. This result is further supported by an evolution-free test that uses only SNe from young, coeval host galaxies across the full redshift range. When the three cosmological probes (SNe, BAO, CMB) are combined, we find a significantly stronger ($>9\sigma$) tension with the $\Lambda$CDM model than that reported in the DESI papers, suggesting a time-varying dark energy equation of state in a currently non-accelerating universe.
\end{abstract}




\section{Introduction} \label{s1}
Recent Baryon Acoustic Oscillation (BAO) measurements from the Dark Energy Spectroscopic Instrument (DESI) project \citep{2025JCAP...02..021A, 2025arXiv250314738D}, when combined with Cosmic Microwave Background (CMB) data from the \citet{2020A&A...641A...5P, 2021A&A...652C...4P} and data release 6 of the Atacama Cosmology Telescope (ACT DR6; \citealt{2024ApJ...962..113M, 2024ApJ...962..112Q, 2024ApJ...966..138M}), exhibit a 3.1$\sigma$ deviation from the $\Lambda$CDM model. Notably, this result does not support the standard model of dark energy, the cosmological constant, but instead favors a time-varying dark energy equation of state. A significant yet not widely recognized outcome drawn from the DESI BAO analysis is that the cosmological parameters obtained by combining BAO and CMB data alone indicate values ($ w_0 = -0.42 $, $ w_a = -1.75 $, $ \Omega_m = 0.353 $; see Table~5 of \citealt{2025arXiv250314738D}) that favor not only a decelerated expansion for the future universe but also suggest a sign of decelerated expansion ($ q_0 = 0.092 \pm 0.20$; see Section~\ref{s5} below) even at the present time, rather than an accelerated expansion. Only when the Type Ia supernova (SN~Ia) dataset \citep{2022ApJ...938..110B, 2024ApJ...973L..14D, 2025ApJ...986..231R} is added to the BAO and CMB measurements does the present universe remain in a state of accelerated expansion, as is widely recognized \citep{1998AJ....116.1009R, 1999ApJ...517..565P}, even though the future universe will transition to a state of decelerated expansion.

However, this result was obtained without incorporating the SN progenitor age-bias effect in the SN dataset. In recent years, direct age measurements of SN host galaxies have revealed a significant correlation between the standardized SN magnitude and progenitor age, which is expected to cause a serious systematic bias with redshift in SN cosmology \citep{2020ApJ...889....8K, 2020ApJ...903...22L, 2022MNRAS.517.2697L}. These results, obtained from host galaxy samples at low redshift ($ z < 0.2 $), have been repeatedly confirmed by other investigators \citep{2021MNRAS.503L..33Z, 2023SCPMA..6629511W}. Paper~I of this series \citep{2025MNRAS.538.3340C} further confirms this age-bias in the SN distance scale at a 5.5$\sigma$ level, based on new age measurements for a larger sample of host galaxies across a broader redshift range ($ z < 0.45 $), illustrating the robustness and universal nature of this systematic bias.

In this paper, in light of the DESI BAO study, we investigate the impact of incorporating the progenitor age-bias effect on SN cosmology when combining the DESI BAO results with the SN dataset. We first examine how the redshift-dependent age-bias alters the distribution of SN data on the Hubble diagram. Subsequently, within the flat-$ w_0w_a $CDM model, we examine the changes that arise after applying the age-bias correction when using the DESI BAO results together with the SN dataset. Finally, when all three cosmological probes (SNe, BAO, CMB) are combined, we present how the cosmological outcomes regarding the accelerated expansion and the cosmological constant would vary depending on whether the age-bias effect is corrected in SN cosmology.

\vspace{-1em}

\section{Robust evidence for strong progenitor age-bias in the supernova distance scale} \label{s2}

\begin{figure*}
\centering
\includegraphics[angle=0,scale=0.186]{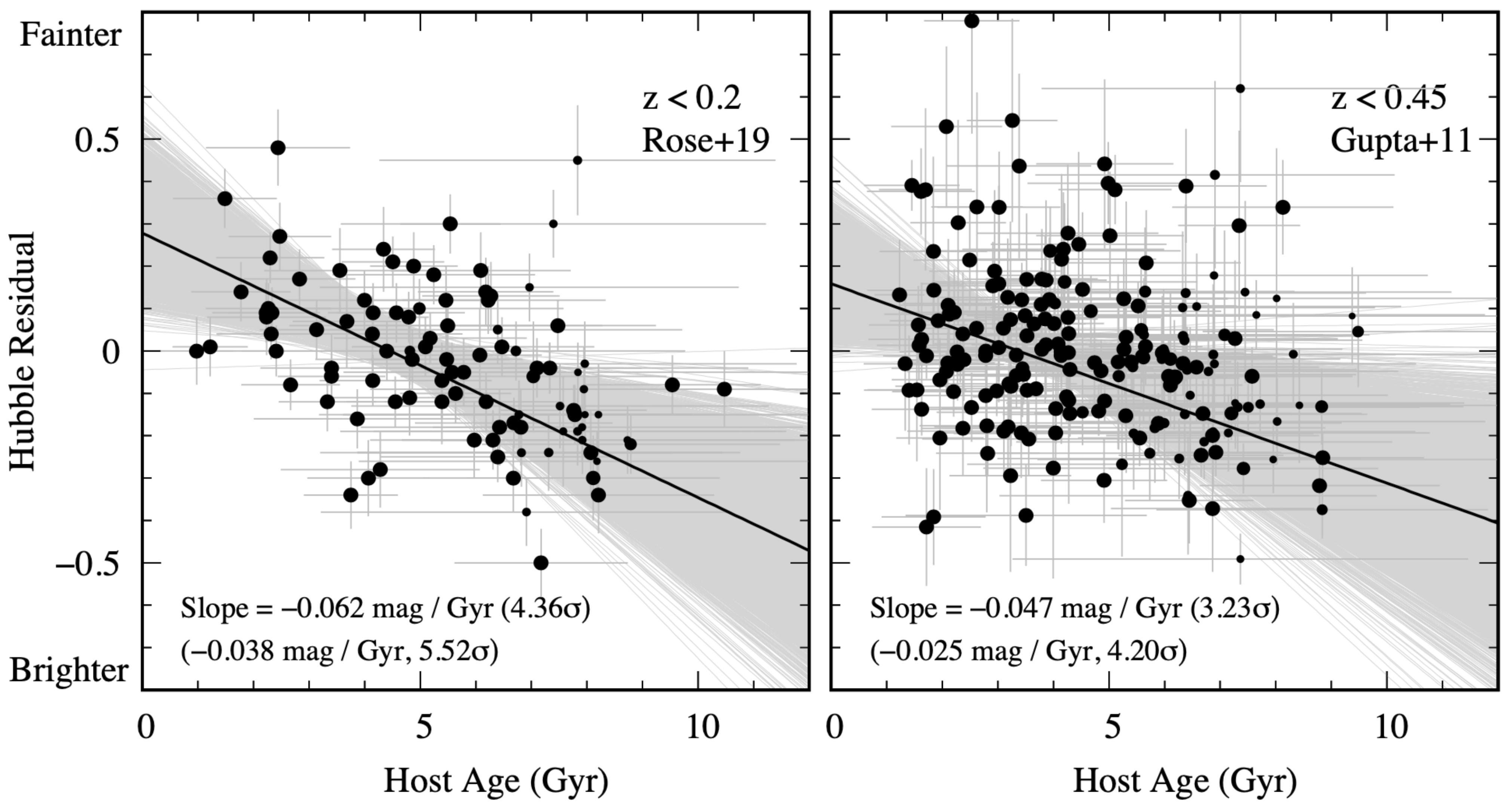}
\caption{
Correlation between population age and HR for SN host galaxies based on our new age measurements reported in Paper~I \citep{2025MNRAS.538.3340C}. The HR is a measure of relative luminosity when the SN sample is confined to a narrow redshift range. The left panel shows the \citetalias{2019ApJ...874...32R} sample, for which the strong significance of the correlation was originally reported by \citet{2020ApJ...903...22L} and has been repeatedly confirmed by third parties at the $>5\sigma$ level \citep{2021MNRAS.503L..33Z, 2023SCPMA..6629511W}. The right panel shows a larger sample ($N \sim 200$) of host galaxies (from \citetalias{2011ApJ...740...92G} sample) over a broader redshift range ($z < 0.45$), confirming the universal nature of the age-bias. The results shown are based on the \texttt{LINMIX} analysis, while the slopes and significances obtained from the full age posteriors are given in parentheses (see \citealt{2025MNRAS.538.3340C}).}
\label{f1}
\end{figure*}

\begin{figure*}
\centering
\includegraphics[angle=0,scale=0.7]{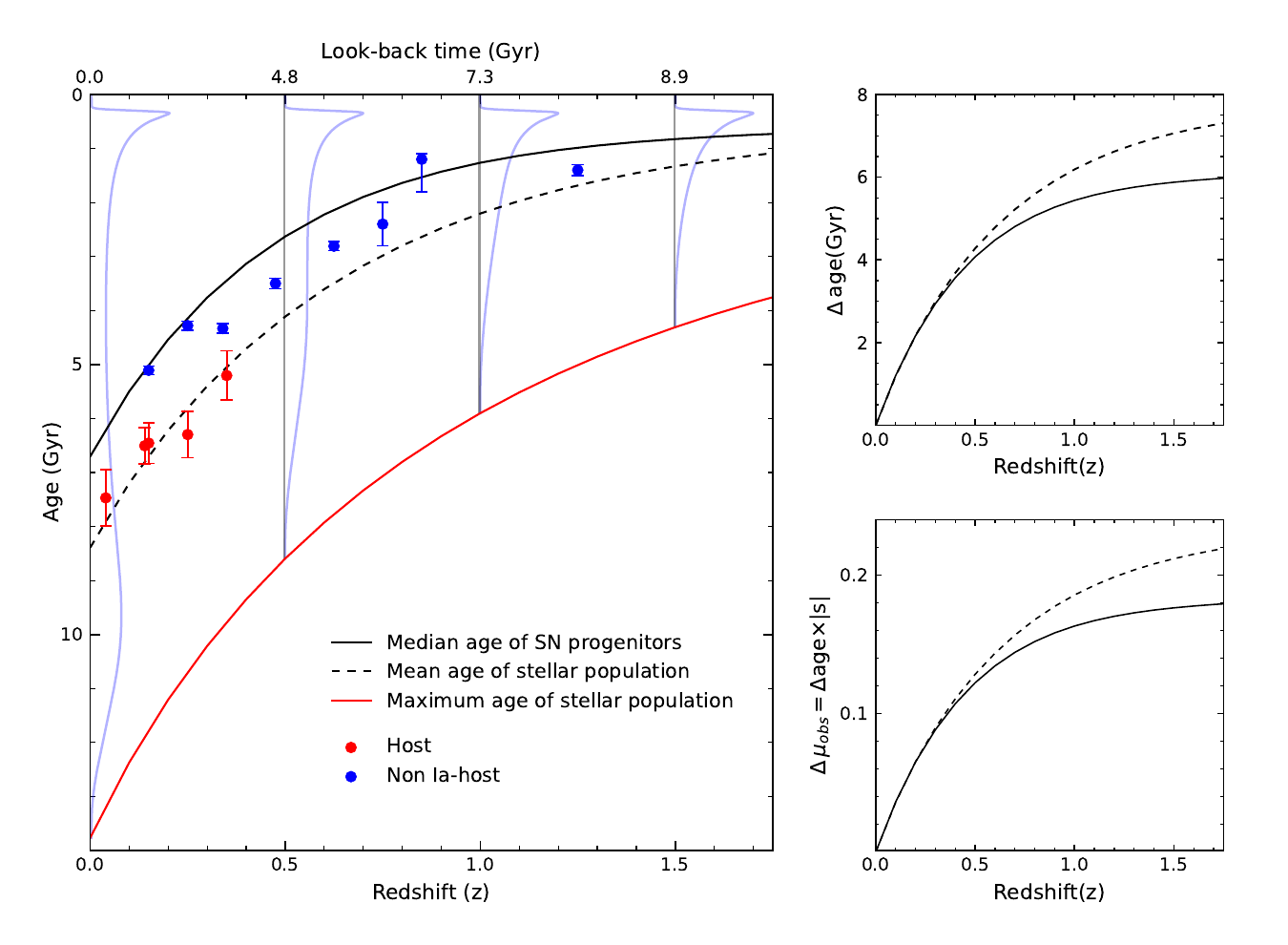}
\caption{
Evolution of stellar population age with redshift. The result is similar to Figure~6 of \citet{2022MNRAS.517.2697L}, but for the $ w_0w_a $CDM model, as suggested by a combined analysis of DESI BAO and CMB \citep{2025arXiv250314738D}. The distribution functions (blue lines) at $z = 0.0$, 0.5, 1.0, and 1.5 represent the SN progenitor age distributions. The black solid line indicates the median age of SN progenitors, while the black dashed line shows the mass-weighted mean age of the stellar population, obtained from the cosmic star formation history and compared with the observed data (red and blue circles). In the right panels, the solid and dashed lines show, respectively, the redshift evolution of the median progenitor age and the mean stellar population age, relative to $z = 0$, and the corresponding variations in Hubble residual, obtained by multiplying the age difference with the age-bias slope $|s|$.}
\label{f2}
\end{figure*}

Many studies reported in the literature have shown that the standardized magnitude of SN~Ia varies with the properties of the host galaxy (\citealt{2010ApJ...715..743K, 2010MNRAS.406..782S, 2014MNRAS.438.1391P, 2020A&A...644A.176R, 2021ApJ...909...26B, 2023MNRAS.520.6214W} and references therein). However, only a few studies have used directly measured population ages of host galaxies to investigate whether the SN progenitor age is the root cause of this variation. \citet{2008ApJ...685..752G}, \citet{2011ApJ...740...92G}, \citet{2013MNRAS.435.1680J}, and \citet{2019ApJ...874...32R} explored this issue using low S/N spectra or multi-band photometric data for host galaxies but did not reach a definitive conclusion. The first notable evidence of an age-bias in the SN distance scale was presented based on high-precision (S/N = 175) spectroscopic observations of early-type host galaxies \citep{2020ApJ...889....8K}. In this study, three different population synthesis models were applied, each producing consistent results showing that the post-standardization magnitudes of SNe from younger host galaxies are systematically fainter than those from older galaxies. However, due to the limited sample size, restricted to nearby normal early-type host galaxies, the statistical significance of these results was at most around 3$\sigma$. Subsequently, \citet{2020ApJ...903...22L} obtained a correlation with much higher statistical significance (4.3$\sigma$), based on the \citet[hear after \citetalias{2019ApJ...874...32R}]{2019ApJ...874...32R} sample of host galaxies encompassing all morphological types. This finding was later confirmed by third-party validations from \cite{2021MNRAS.503L..33Z} and \citet{2023SCPMA..6629511W}. Recently, in Paper~I of this series \citep{2025MNRAS.538.3340C}, we performed new and consistent measurements of population ages for the \citetalias{2019ApJ...874...32R} and \citet[hear after \citetalias{2011ApJ...740...92G}]{2011ApJ...740...92G} host galaxy samples using the latest version of the population synthesis model from \citet{2010ApJ...712..833C}. This analysis further revealed a statistically significant correlation of up to 5.5$\sigma$ between host age and Hubble residual (HR), underscoring that the age-bias is a robust and ubiquitous effect that should be considered a critical systematic bias in SN cosmology (see Figure~\ref{f1}).

The origin of this correlation between progenitor age and HR has been explained by \citet{2022MNRAS.517.2697L}, who has shown that, unlike the key assumption of SN cosmology \citep{2019NatAs...3..706J}, the \citet{1993ApJ...413L.105P} relation (width–luminosity relation, WLR) and the color–luminosity relation (CLR) in the SN luminosity standardization process do strongly depend on progenitor age at the $4.6\sigma$ level. They also found that two subgroups based on other host properties—such as stellar mass, metallicity, and dust content—exhibit only insignificant ($0.98-1.25\sigma$) offsets in the WLR and CLR, suggesting that progenitor age is most likely the root cause of the reported correlations between host properties and HR. This result is based on the \citetalias{2019ApJ...874...32R} sample at relatively low redshift ($z < 0.20$), but a comparable result is also obtained from the \citetalias{2011ApJ...740...92G} sample at higher redshift ($z < 0.45$), further illustrating the universal nature of this age-bias (Park et al. 2025, in prep.). \citet{2023ApJ...959...94C} also reported that the root cause of the host mass step \citep{2010ApJ...715..743K, 2010MNRAS.406..782S}, the property most widely included in SN~Ia luminosity standardization, is the stellar population age, based on the empirical galaxy color–magnitude relation. A more detailed analysis focused on how the linear correlation between progenitor age and HR leads to the mass step and the star formation rate step \citep{2020A&A...644A.176R} in SN~Ia magnitudes will be presented in Paper~III of this series (Park et al.\ 2025, in prep.). It is important to note that the luminosity corrections based on the commonly considered host mass step cannot account for the progenitor age-bias effect (see Section~\ref{s43} below). This is because host mass and age evolve very differently with redshift. Within the redshift range most relevant to SN cosmology (0.0~$< z <$~1.0), the evolution of host mass is relatively small \citep{2004ApJ...606.1029B, 2007MNRAS.379.1491C}, whereas the evolution of host age is significant—amounting to 5–6~Gyr 
\citep{2014MNRAS.445.1898C, 2022MNRAS.517.2697L}. However, the effect of progenitor age-bias is still not adequately accounted for in most SN cosmology studies.

In the following section, we investigate how and to what extent the progenitor age-bias impacts SN cosmology analyses. To do so, we use the average value of the age-bias slopes from previously reported results across three different samples. For the \citet{2020ApJ...889....8K}, the weighted mean of the slopes obtained from three population synthesis models is $ s = -0.025 \pm0.006$~mag~Gyr$^{-1} $. For the \citetalias{2019ApJ...874...32R} and \citetalias{2011ApJ...740...92G} samples, the new results obtained by \citet{2025MNRAS.538.3340C} are $ s = -0.038\pm0.007 $ and $ s = -0.025\pm0.006 $~mag~Gyr$^{-1} $, respectively (see Figure~\ref{f1}). We adopt the average value, $ -0.030 \pm 0.004 $~mag~Gyr$^{-1} $, of these three results from different host galaxy samples at different redshift bins. 

\section{Redshift evolution of progenitor age and its cosmological impact}\label{s3}

\begin{figure*}
\centering
\includegraphics[angle=0,scale=0.4]{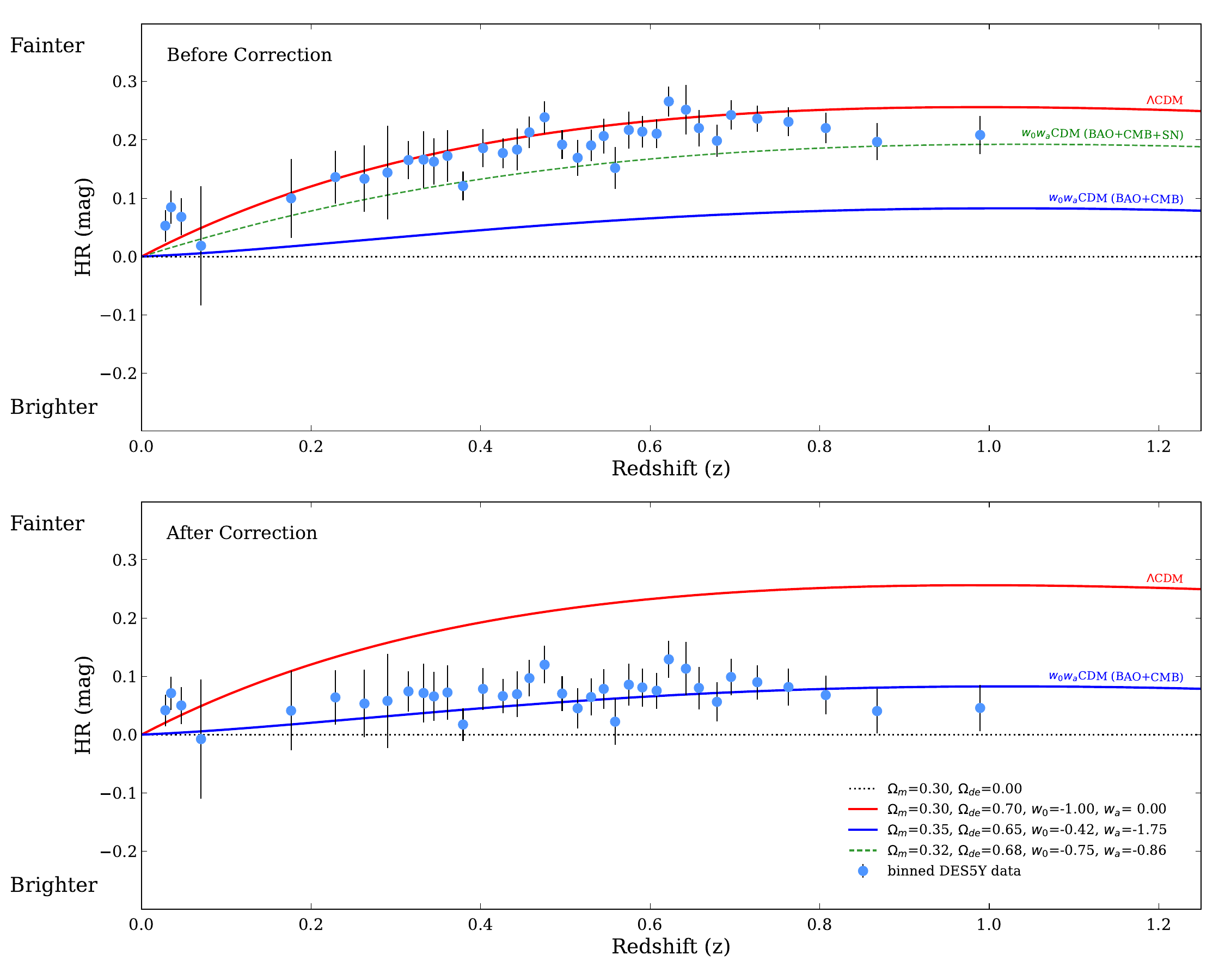}
\caption{
The residual Hubble diagram before (top panel) and after (bottom panel) the age-bias correction. The corrections are applied to the observational data from the DES SN project \citep{2024ApJ...973L..14D}, using the $\Delta\mathrm{HR}/\Delta\mathrm{age}$ slope ($0.030\pm0.004$~mag~Gyr$^{-1}$) and the redshift evolution of the median progenitor age relative to $z = 0.0$, as shown in Figure~\ref{f2}. After the correction, the SN dataset no longer supports the $\Lambda$CDM model (red solid lines), but instead shows better consistency with a time-varying dark energy equation of state, as described by the flat-$ w_0w_a $CDM model favored by the combined BAO and CMB analysis (blue solid lines). As shown in the top panel, this model from BAO+CMB alone deviates significantly from that based on the combined analysis of BAO, CMB, and uncorrected SN data (green dashed line). In both cases, the relevant cosmological parameters are adopted from \citet{2025arXiv250314738D}. A binning scheme similar to that of the DES SN project was implemented to ensure $\sim$50 SNe per each bin. In the bottom panel, an uncertainty in the age-bias slope was propagated into the total error budget of the binned data. A similar trend is also obtained when using the Pantheon+ dataset \citep{2022ApJ...938..110B}.}
\label{f3}
\end{figure*}

\begin{figure*}
\centering
\includegraphics[angle=0,scale=0.4]{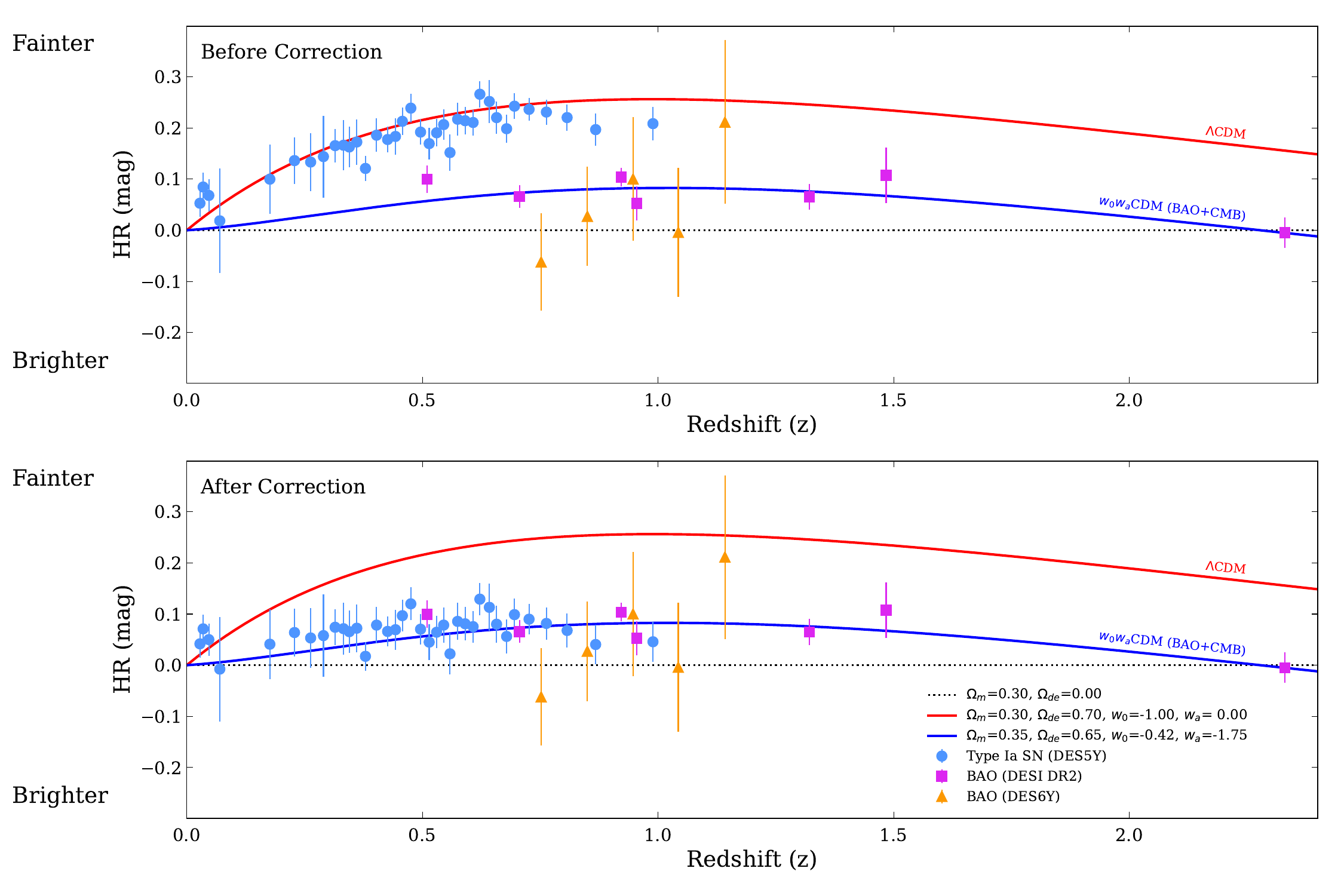}
\caption{
Similar to Figure~\ref{f3}, but showing the comparison between SN and BAO distance measurements. The BAO distances are taken from \citet[DESI DR2 shown in magenta squares]{2025arXiv250314738D} and \citet[DES6Y shown in orange triangles] {2025arXiv250306712D}, and have been converted to distance moduli. After correcting for the progenitor age-bias (bottom panel), the SN and BAO distance scales show good agreement. For a self-consistent comparison, we adopt the value of the product $H_0 \cdot r_d$ from the combined DESI BAO (DR2) and CMB analysis within the flat-$ w_0w_a $CDM model (see text for details).}
\label{f4}
\end{figure*}

The progenitor age-bias in the SN distance scale, as described in the previous section, propagates into a strong redshift-dependent bias in SN magnitudes due to the redshift evolution of progenitor age. This connection was already acknowledged in the original discovery papers \citep{1998AJ....116.1009R, 1998ApJ...507...46S, 1999ApJ...517..565P}. For example, \citet{1998AJ....116.1009R} stated, \textit{“we can place empirical constraints on the effect that a change in the progenitor age would have on our SN~Ia distances by comparing subsamples of low-redshift SNe~Ia believed to arise from old and young progenitors.”} Therefore, part of the variation in HRs with increasing redshift, as observed in the residual Hubble diagram, results from the redshift evolution of progenitor age.\footnote{For other possible interpretations of the relative dimming of SNe at high redshift, the readers are referred to \citet{2000ApJ...532...28A}, \citet{2004MNRAS.350..729I}, and \citet{2022IJMPD..3150104L}.} Accordingly, a correction is required to remove this effect in order to isolate the purely cosmological signal. The most direct approach for correcting this effect would be to measure the ages of all host galaxies used in cosmological analyses across the full redshift range and apply a relative age-bias correction to the SN distance modulus on a galaxy-by-galaxy basis. However, currently available host galaxy age measurements exist only for a limited number of samples at $ z < 0.45 $ \citep{2019ApJ...874...32R, 2020ApJ...889....8K, 2025MNRAS.538.3340C}, and therefore, such direct corrections cannot yet be applied to all host galaxies. Nevertheless, the evolution of the mean stellar population age within galaxies as a function of redshift can be inferred from the well-established cosmic star formation history (CSFH). The SN~Ia progenitor age distribution (SPAD) can also be reliably obtained using the empirically derived SN~Ia delay-time distribution (DTD) in conjunction with the CSFH \citep{2014MNRAS.445.1898C}. 

In Figure~\ref{f2}, following \citet{2020ApJ...889....8K} and \citet{2022MNRAS.517.2697L}, we adopt this method to estimate the redshift evolution of the SPAD. As shown by \citet{2014MNRAS.445.1898C}, this result is not strongly affected by current uncertainties in the DTD. At low redshift, the SPAD shows a distinct bimodality, with young (prompt) and old (delayed) components. At higher redshift, however, the distribution becomes more uniform, dominated by the young population. Therefore, SN progenitors in host galaxies become progressively younger, on average, with increasing redshift, showing a similar trend to the evolution of the mean stellar population age derived from the CSFH. This prediction is well supported by directly measured ages from observational data for both SN~Ia host galaxies and non-Ia host galaxies at present. Note that SN~Ia arises in all types of galaxies and, thus, all galaxies were once SN~Ia host galaxies. Therefore, SN~Ia host galaxies in the sample used in the cosmological analysis—if we had measured population ages for all of them—would also follow the same redshift evolution of age predicted in Figure~\ref{f2}. Assuming a linear relationship between host age and HR, we can thus apply a mean relative correction to the observed SN magnitude as a function of redshift, relative to $ z = 0.0 $. This relative average correction is obtained by multiplying the relative age difference by the slope of the age-bias derived in Section~~\ref{s2}:
\[
\Delta m(z) = \Delta \mathrm{age}(z) \times 0.030~\mathrm{mag~Gyr}^{-1}.
\tag{1}
\]

As illustrated in the right panels of Figure~\ref{f2}, within the redshift range most relevant to SN cosmology ($0 < z < 1$), we expect, on average, a 5.3~Gyr variation in progenitor age (top panel) and, therefore, a $\sim 0.16$~mag variation in SN luminosity (bottom panel). This relative correction, $\Delta m(z)$, with respect to $z = 0.0$, is then integrated into the formula for the standardized distance modulus from SNe~Ia
\[
\mu_{\mathrm{SN}} = m - M + \alpha x_1 - \beta c - \Delta m(z),
\tag{2}
\]
where $ m $ is the apparent magnitude, $ M $ is the absolute magnitude at the reference point $(x_1 = 0.0,\, c = 0.0)$, $ x_1 $ and $ c $ are the light curve width and color parameters, and $ \alpha $ and $ \beta $ are the absolute values of the slopes of the WLR and CLR, respectively. For the values of $\alpha$ and $\beta$, we adopt 0.148 and 3.09 for the Pantheon+ sample \citep{2022ApJ...938..110B}, and 0.161 and 3.12 for the DES-SN5YR data \citep{2024ApJ...973L..14D}. We apply this average correction to all SN data at each redshift. Although this approach does not reduce the scatter in HR values caused by variation in progenitor age at a given redshift, it does allow for an appropriate correction of the mean HR value in each redshift bin. Since cosmological parameters are mostly determined based on the average values of SN data within each redshift bin, applying a mean correction still allows us to predict the resulting changes in the cosmological model. While the result in Figure~\ref{f2} is derived under the flat-$ w_0w_a $CDM model using the cosmological parameters predicted from the DESI BAO (DR2) combined with CMB data (Table~5 of  \citealt{2025arXiv250314738D}), our corrections for the progenitor age-bias in the cosmological analysis below are self-consistently determined with the cosmological fit.

Figure~\ref{f3} compares the binned SN data from the Dark Energy Survey (DES) project \citep{2024ApJ...973L..14D} with cosmological models in the residual Hubble diagram, where $\mathrm{HR} = \mu_{\mathrm{SN}} - \mu_{\mathrm{model}}$. Following \citet{2024ApJ...973L..14D}, the $H_0$ and the SN~Ia absolute magnitude $M$ are combined in the single parameter $M + 5{\log}(c/H_0) = -11.80$. Therefore, the value of $H_0$ has no impact on the cosmological analysis. Before the age-bias correction, the SN data are broadly consistent with the $\Lambda$CDM model. After correcting for the age-bias, however, the SN dataset no longer supports the $\Lambda$CDM model. Instead, interestingly, it aligns more closely with a time-varying dark energy equation-of-state model (flat-$ w_0w_a $CDM), as recently suggested by the DESI BAO (DR2) project in a combined analysis of BAO and CMB \citep{2025arXiv250314738D}. As compared in the top panel of Figure~\ref{f3}, it is important to note that this model derived from BAO+CMB alone (the blue line) is even further away from the $\Lambda$CDM model than the $w_0w_a$CDM model most extensively discussed in the DESI papers, which is based on the combined analysis of BAO, CMB, and SNe before age-bias correction (the green dashed line). 

Within the same $ w_0w_a $CDM model favored by BAO+CMB, in Figure~\ref{f4}, comparison is made between the SN and BAO distances in the residual Hubble diagram. The employed BAO distances are transverse comoving distances, $ d_M(z) $, adopted from \citet{2025arXiv250314738D}. Additional BAO distance data are also adopted from \citet {2025arXiv250306712D}. To directly compare with SN distances, these BAO distances are transformed to the luminosity distance via $ d_L(z) = (1 + z) d_M(z) $. In order to be fully internally consistent, the value for the product of $ H_0 $ and the sound horizon ($ r_d $) is also adopted from the same combined analysis of DESI BAO (DR2) and CMB in the $ w_0w_a $CDM model (See Section~\ref{s43}; see also Table~7 of \citealt{2025arXiv250314738D}). Within this fixed value $(r_d h = 93.9 \pm 2.8\ \rm{Mpc})$, any combination of $ H_0 $ and $ r_d $ produces the same result. If the $\Lambda$CDM model had been assumed, the value of $r_d h$ would increase and the BAO distances would shift closer to the $\Lambda$CDM prediction. However, given that the DESI project favors the $w_0w_a$CDM model, we consider it appropriate to adopt the $r_d h$ value derived from this model. It is clear from Figure~\ref{f4} that, before the age-bias correction, the SN distances do not match with the BAO distances. After the correction, however, the SN distance scale matches very well with the BAO distance scale in the $ w_0w_a $CDM model favored by BAO+CMB. Therefore, the revised standard candle (SNe Ia) is in agreement with the standard ruler (BAO). For more detailed analysis and cosmological parameter estimation, following the DESI BAO papers \citep{2025JCAP...02..021A, 2025arXiv250314738D}, we have combined SN dataset with BAO and CMB data in the flat-$ w_0w_a $CDM model, and compared that with the probability distribution for BAO combined only with CMB. The result of this analysis is presented in the following section.

\section{Cosmological analysis} \label{s4}

\subsection{Datasets and Cosmological Calculations}

In this paper, we employ two different SN~Ia datasets, the Pantheon+ sample \citep{2022ApJ...938..110B} and the DES-SN5YR Data Release \citep{2024ApJ...973L..14D}. For the BAO measurements, we make use of the DESI BAO (DR2) dataset \citep{2025arXiv250314738D}. For the CMB analysis, multiple datasets are combined following the methodology described in the DESI DR2 paper. 
These include the \textit{SimAll} and \textit{Commander} likelihoods from the Planck PR3 release~\citep{2020A&A...641A...5P}, the \textit{CamSpec} likelihood based on the Planck PR4 NPIPE maps~\citep{2022MNRAS.517.4620R}, and the CMB lensing measurements from the Planck PR4 NPIPE release~\citep{2022JCAP...09..039C}, together with the ACT DR6 \citep{2024ApJ...962..113M, 2024ApJ...962..112Q, 2024ApJ...966..138M}. In the following sections, we refer to the DES-SN5YR dataset as \textit{DES5Y}, the Pantheon+ dataset as \textit{Pantheon+} in the main text but abbreviated as "\textit{Panth.}" in tables and figures, the DESI DR2 dataset as \textit{BAO}, and the combined Planck+ACT dataset as \textit{CMB}.

We consider two cosmological models under the assumption of flat curvature: the $w$CDM model (see Section~\ref{s42}), where the dark energy equation-of-state parameter $w$ is constant, and the $w_0w_a$CDM model (see Section~\ref{s43}), parameterized via the Chevallier-Polarski-Linder (CPL; \citealt{2001IJMPD..10..213C, 2003PhRvL..90i1301L}) form, in which $w$ evolves with time. Bayesian analysis of the model parameters is performed via MCMC sampling \citep{2002PhRvD..66j3511L, 2013PhRvD..87j3529L, 2005math......2099N} using \textsc{Cobaya} \citep{2021JCAP...05..057T}, while cosmological calculations are carried out using the Boltzmann code \textsc{CAMB} \citep{2000ApJ...538..473L, 2012JCAP...04..027H}. To evaluate the impact of the age-bias correction on the findings of \citet{2025arXiv250314738D}, we set our free parameters and priors as closely as possible to those used in their paper. For the analysis based solely on SN data, for consistency, we also adopt the same priors on \(\Omega_m\), \(w_0\), and \(w_a\) as in \citet{2025arXiv250314738D}. As described in Section~\ref{s3}, SN-only result is not affected by the value of \(H_0\) \citep[see][]{2024ApJ...973L..14D}. As the age-bias correction is applied exclusively to the SN~Ia data, our focus remains on the parameters $\Omega_m$, $w_0$, and $w_a$, which are most directly constrained by SN~Ia observations. We also report the present-day deceleration parameter, $q_0$, derived from $w_0$ and $\Omega_m$, where a negative $q_0$ indicates an accelerating universe and a positive $q_0$ corresponds to a decelerating universe. Thus, we present four key parameters: $\Omega_m$, $w_0$, $w_a$, and $q_0$. Parameter estimates, obtained with \textsc{GetDist}~\citep{2019arXiv191013970L}, are presented as means with standard deviations for symmetric distributions or as means with 68\% credible intervals in other cases. 

\subsection{Flat wCDM model} \label{s42}

\begin{figure*}
\includegraphics[angle=0,scale=0.1]{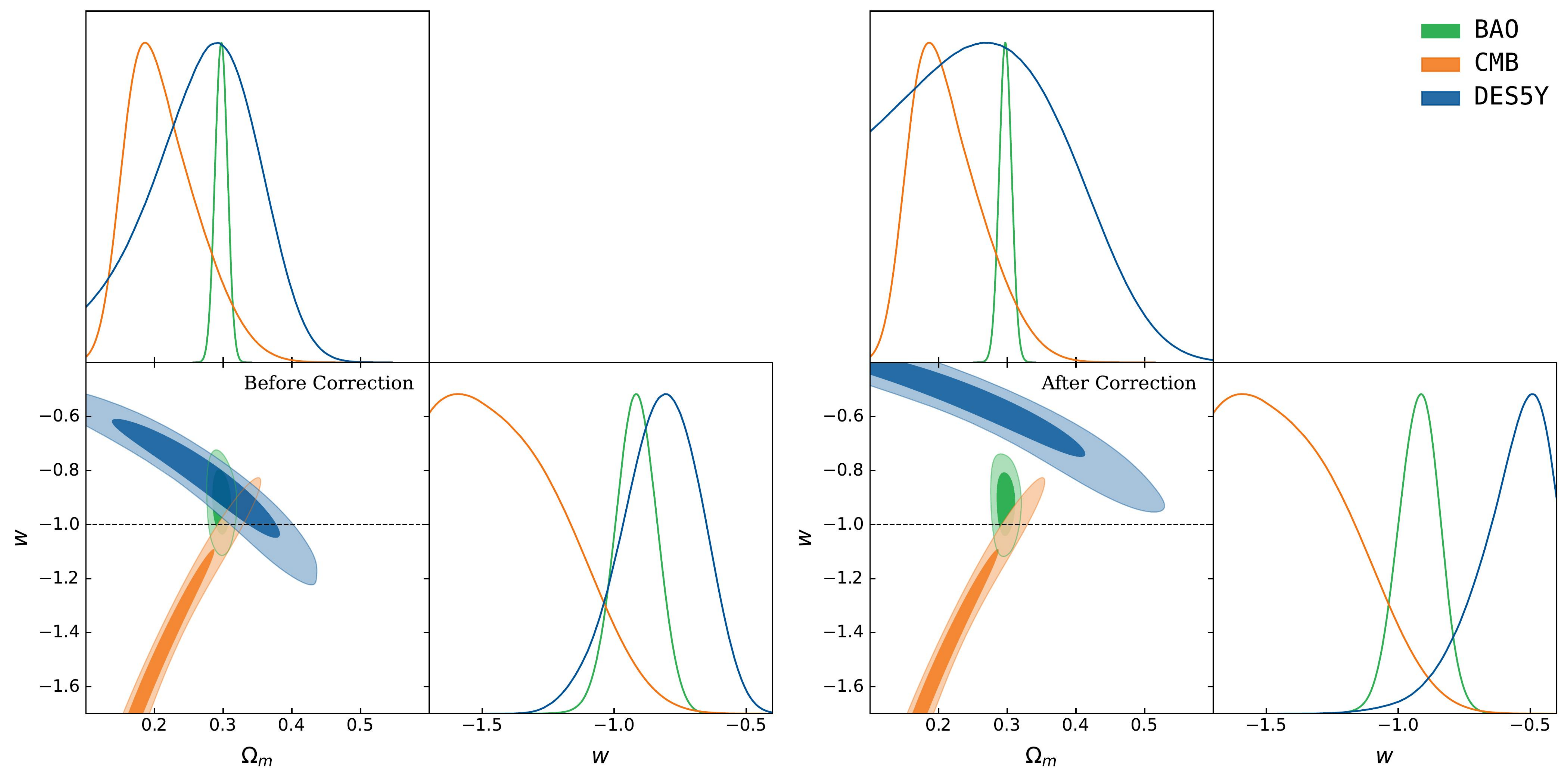}
\caption{
Confidence contours (68$\%$ and 95$\%$) in the $\Omega_m - w$ plane for the flat-$w$CDM model. The DES5Y SN data, shown before and after the age-bias correction (left and right panels, respectively), is compared with BAO and CMB data. In the left panel, a small overlap near $w=-1$ weakly supports the $\Lambda$CDM model when all three datasets are combined. In the right panel, after applying the age-bias correction, the SN contours shift upward, favoring a significantly larger value of $w$ ($\sim-0.56$). The three probes no longer intersect near $w=-1$, but instead show a progression in the mean $w$ values: from the CMB ($w \approx -1.5, z\approx 1090$), to BAO ($w \sim -0.92, 0.4 < z < 4$), to SNe Ia ($w \sim -0.56, 0 < z < 1.5$), suggesting a redshift evolution of the dark energy equation-of-state parameter.
\label{f5}}
\end{figure*}

\renewcommand{\arraystretch}{1.5} 
\begin{table*}
\caption{Cosmological parameter constraints in the flat-$ w $CDM model. Parameter estimates are presented as means with standard deviations for approximately symmetric posterior distributions, or as means with 68\% credible intervals for non-Gaussian cases. Results are shown for individual datasets—BAO only, CMB only, and SN only—both before and after the age-bias correction applied to the SN data.}
\label{tab:cosmo_params_1}
    \hspace{0pt}
    \begin{tabular}{lcccccc}
        \toprule
        \textbf{dataset}               & $\Omega_m$                 & $w$  & $q_0$\\
        \midrule BAO & $0.297 \pm 0.009$ & $-0.917 \pm 0.078$ & $-0.467 \pm 0.082$ \\
CMB & $0.211^{+0.023}_{-0.068}$ & $-1.491 \pm 0.272$ & $-1.285 \pm 0.426$ \\
Panth. & $0.277 \pm 0.073$ & $-0.894 \pm 0.148$ & $-0.455 \pm 0.071$ \\
DES5Y & $0.265 \pm 0.083$ & $-0.821 \pm 0.149$ & $-0.388 \pm 0.071$ \\
Panth. (corrected) & $0.252 \pm 0.117$ & $-0.595^{+0.176}_{-0.094}$ & $-0.143^{+0.064}_{-0.044}$ \\
DES5Y (corrected) & $0.247 \pm 0.125$ & $-0.556^{+0.181}_{-0.078}$ & $-0.102^{+0.066}_{-0.042}$ \\ 
        \bottomrule
    \end{tabular}
\end{table*}

The $w$CDM model extends the $\Lambda$CDM model by introducing the dark energy equation-of-state parameter $w$, which is assumed to be constant. In this subsection, we examine how the parameter estimates in the flat-$w$CDM model are affected by applying the age-bias correction to the SN data. Figure~\ref{f5} shows constraints on $\Omega_m$ and $w$ for the BAO, CMB, and DES5Y datasets. The left and right panels correspond to DES5Y data before and after the age-bias correction, respectively. Before the age-bias correction, the SN data yield $w = -0.82 \pm 0.15$. The CMB contours overlap only marginally with those from SNe and BAO, but a small overlapping region near $w=-1$ still supports the cosmological constant when all three probes are combined.

On the other hand, after applying the age-bias correction to the DES5Y data, the value of $w$ increases to $w = -0.56 ^{+0.18}_{-0.08}$. Now the SN contours deviate from those of the BAO and CMB by more than $2\sigma$ in the $w$–$\Omega_m$ plane, breaking the previous concordance and revealing significant tension among the datasets. Table~\ref{tab:cosmo_params_1} presents the estimated values of $\Omega_m$, $w$, and $q_0$ for the BAO, CMB, and SN datasets, including the Pantheon+. As shown in the table, the Pantheon+ exhibit a similar increase in $w$ after the age-bias correction. An interesting trend is observed among three probes at different cosmic epochs. The CMB, a high-redshift ($z \sim 1090$) probe, yields a lowest $w$ value ($w = -1.5$), whereas the BAO, an intermediate-redshift ($0.4 < z \lesssim 4$) probe, tends to produce an intermediate $w$ value ($w = -0.92$). After the age-bias correction, the SN data, covering the lowest redshift range ($0 < z \lesssim 1.5$), exhibit $w \sim -0.6$, a highest $w$ value among all probes considered. The $w$ values inferred from different probes exhibit a monotonic increase with decreasing redshift, suggesting a redshift-dependent evolution of the dark energy equation of state. This observed trend provides a compelling motivation to investigate the dynamical dark energy model in the following subsection.

\begin{figure*}
\includegraphics[angle=0,scale=0.5]{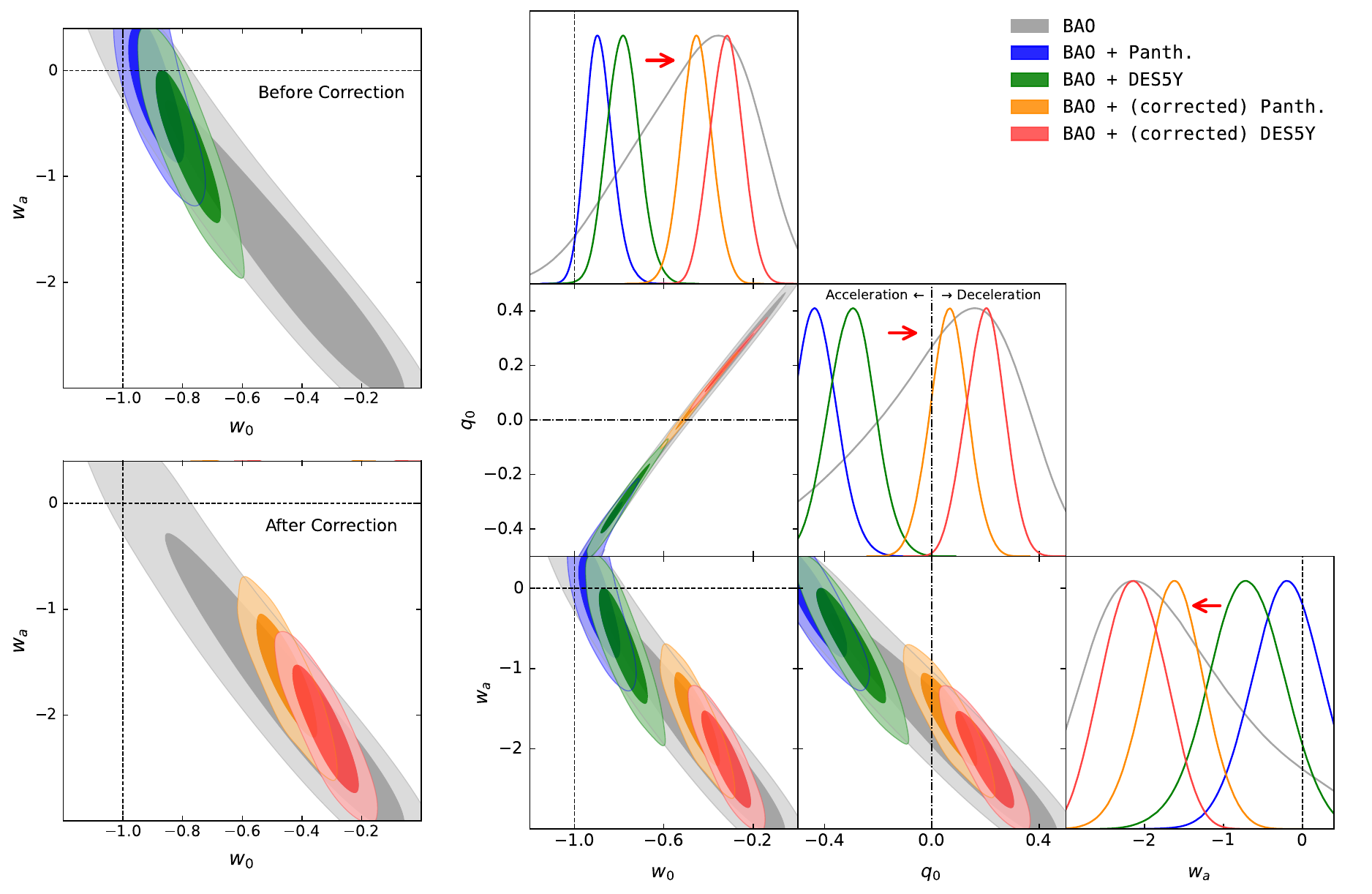}
\caption{
Posterior distributions in the $w_0-w_a$,  $w_0-q_0$ and $q_0-w_a$ parameter spaces for the flat-$ w_0 w_a $CDM model. The contours represent 68\% and 95\% credible intervals, derived from BAO alone (gray) and BAO combined with Pantheon+ or DES5Y SN datasets before (blue and green) and after (orange and red) the age-bias correction. After applying the age-bias correction, the combined (BAO+SNe) constraints become more consistent with the BAO-only results.
\label{f6}}
\end{figure*}

\renewcommand{\arraystretch}{1.5} 
\begin{table*}
\caption{Cosmological parameter constraints in the flat-$ w_0w_a $CDM model. As in Table~1, parameter estimates are presented either as means with standard deviations or as means with 68\% credible intervals. The results are distinguished by whether or not CMB data are included. For each combination, results are shown both before and after applying the age-bias correction to the SN data, with the corrected cases explicitly labeled.}
\label{tab:cosmo_params_2}
    \hspace{5pt}
    \begin{tabular}{lcccccc}
        \toprule
        \textbf{dataset}               & $\Omega_m$                & $w_0$                   & $w_a$ & $q_0$\\
        \midrule BAO & $0.353^{+0.040}_{-0.020}$ & $-0.473^{+0.336}_{-0.184}$ & $-1.687^{+0.463}_{-1.234}$ & $0.028^{+0.361}_{-0.168}$ \\
Panth. & $0.319 \pm 0.095$ & $-0.923 \pm 0.147$ & $-0.634 \pm 1.015$ & $-0.428 \pm 0.102$ \\
DES5Y & $0.376^{+0.069}_{-0.025}$ & $-0.820 \pm 0.131$ & $-1.771^{+0.425}_{-1.206}$ & $-0.261^{+0.118}_{-0.096}$ \\
Panth. (corrected) & $0.360 \pm 0.127$ & $-0.652^{+0.187}_{-0.132}$ & $-0.842 \pm 0.908$ & $-0.104 \pm 0.091$ \\
DES5Y (corrected) & $0.427^{+0.103}_{-0.035}$ & $-0.562^{+0.151}_{-0.120}$ & $-1.748^{+0.472}_{-1.216}$ & $0.028 \pm 0.096$ \\
BAO + Panth. & $0.301^{+0.022}_{-0.011}$ & $-0.889 \pm 0.060$ & $-0.196 \pm 0.437$ & $-0.433 \pm 0.078$ \\
BAO + DES5Y & $0.319^{+0.017}_{-0.011}$ & $-0.782 \pm 0.071$ & $-0.713 \pm 0.474$ & $-0.300 \pm 0.086$ \\
BAO + Panth. (corrected) & $0.359 \pm 0.011$ & $-0.453 \pm 0.068$ & $-1.633 \pm 0.381$ & $0.064 \pm 0.070$ \\
BAO + DES5Y (corrected) & $0.374 \pm 0.011$ & $-0.320 \pm 0.072$ & $-2.136 \pm 0.386$ & $0.199 \pm 0.070$ \\
\midrule
\multicolumn{5}{l}{\textbf{CMB data included}} \\BAO + CMB & $0.352 \pm 0.021$ & $-0.430 \pm 0.207$ & $-1.704 \pm 0.580$ & $0.075 \pm 0.215$ \\
BAO + CMB + Panth. & $0.311 \pm 0.006$ & $-0.838 \pm 0.054$ & $-0.612 \pm 0.201$ & $-0.366 \pm 0.061$ \\
BAO + CMB + DES5Y & $0.319 \pm 0.006$ & $-0.754 \pm 0.056$ & $-0.848 \pm 0.217$ & $-0.271 \pm 0.061$ \\
BAO + CMB + Panth. (corrected) & $0.353 \pm 0.006$ & $-0.447 \pm 0.059$ & $-1.585 \pm 0.227$ & $0.065 \pm 0.060$ \\
BAO + CMB + DES5Y (corrected) & $0.363 \pm 0.006$ & $-0.337 \pm 0.062$ & $-1.902 \pm 0.246$ & $0.178 \pm 0.061$ \\
        \bottomrule
    \end{tabular}
\end{table*}

\begin{figure*}
\includegraphics[angle=0,scale=0.5]{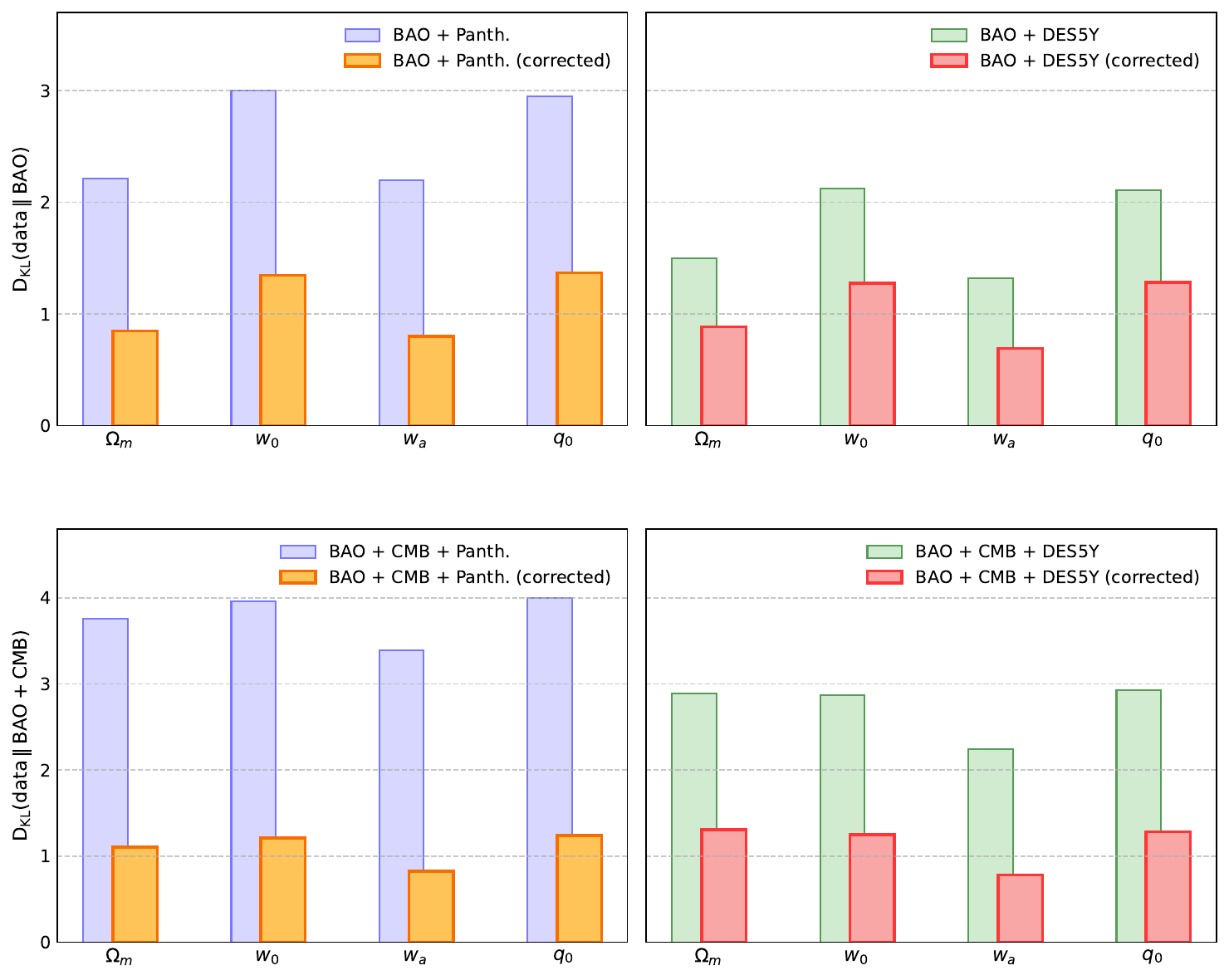}
\caption{
Results of the Kullback–Leibler (KL) divergence analysis. The SN datasets are Pantheon+ (left) and DES5Y (right). The upper panels show the KL divergence of marginalized cosmological parameter posteriors, using BAO data alone as the reference. In both cases, the BAO+SNe combinations exhibit lower KL divergence after the age-bias correction, indicating improved consistency with the BAO-only results. The lower panels display the KL divergence relative to the BAO+CMB combination. Again, the application of the age-bias correction consistently reduces the KL divergence for all parameters, indicating improved agreement between the SN-added results (BAO+CMB+SN) and the BAO+CMB.
\label{f7}}
\end{figure*}

\subsection{Flat $w_0w_a$CDM model: A new concordance} \label{s43}

\begin{figure*}
\centering
\includegraphics[angle=0,scale=0.5]{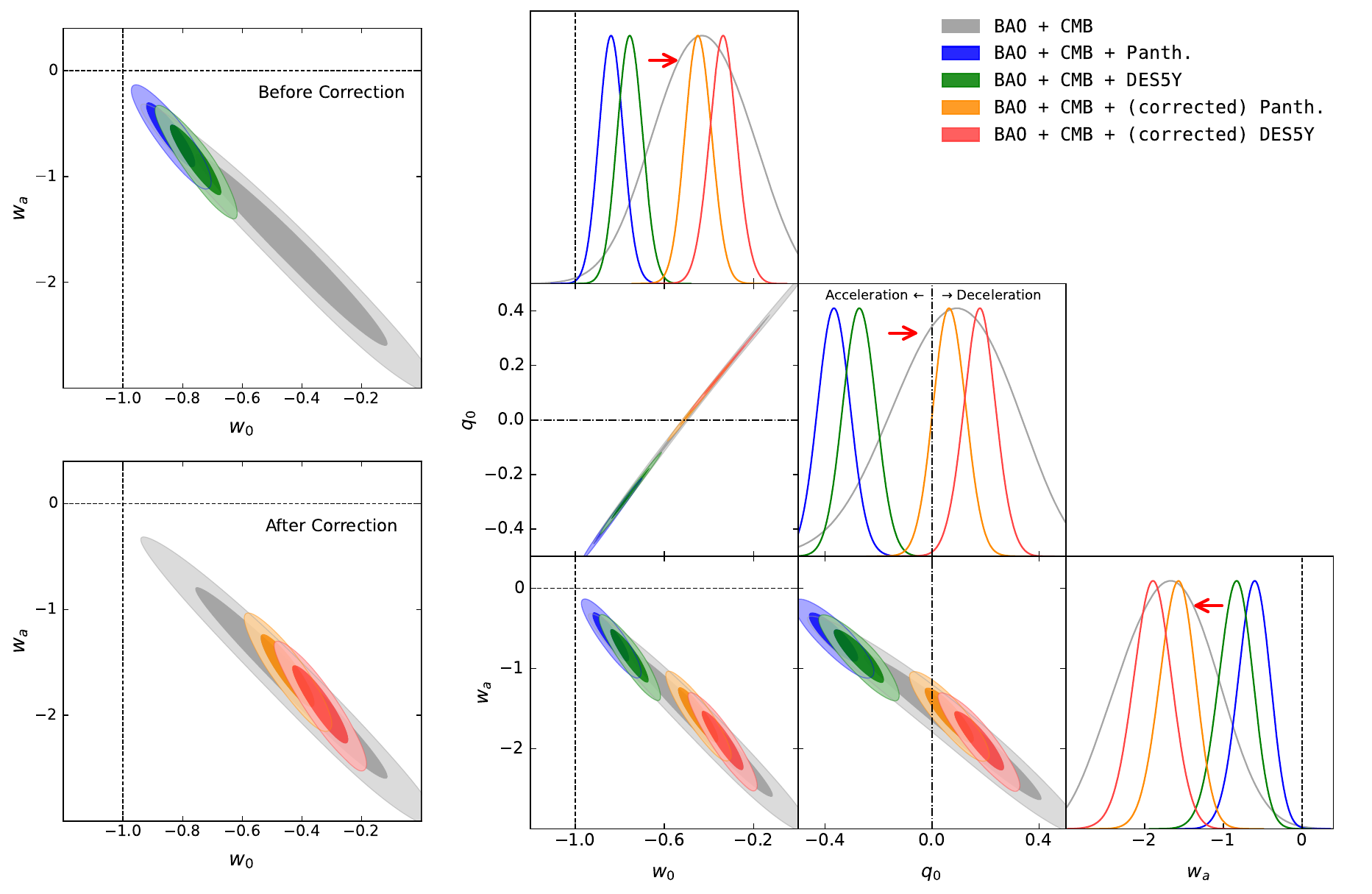}
\caption{
Same as Figure~\ref{f6}, but with CMB data included in the analysis. The contours are derived from the combination of BAO+CMB (gray), and BAO+CMB combined with each of the Pantheon+ and DES5Y SN~Ia datasets before (blue and green) and after (orange and red) the age-bias correction, respectively. After applying the age-bias correction, the SN-added results (BAO+CMB+SNe) align more closely with the BAO+CMB posteriors across all three parameters ($w_0, w_a, q_0$).}
\label{f8}
\end{figure*}

The CPL parameterization, expressed as $w(a) = w_0 + w_a(1 - a)$, provides the simplest framework for describing dynamical dark energy. Here, $w_0$ represents the present-day value of the dark energy equation of state, while $w_a$ quantifies its variation with respect to the scale factor. Motivated by the findings of the \citet{2025JCAP...02..021A} and \citet{2025arXiv250314738D}, we analyze the behavior of the constrained parameters $w_0$, $w_a$, and $q_0$ within the CPL parameterization when the age-bias correction is applied to SN~Ia data. 

Figure~\ref{f6} illustrates the effect of age-bias correction on the two SN datasets in combination with the BAO data. The BAO-only contour (gray) provides a reference for comparing the shift in the BAO+SNe contours before and after age-bias correction. The BAO data alone yield $w_0 = -0.47^{+0.34}_{-0.18}$, and $w_a = -1.69^{+0.46}_{-1.23}$, indicating a mild deviation from the $\Lambda$CDM values, consistent with \citet{2025arXiv250314738D}. The mean value of $q_0$ based on the BAO data is greater than zero ($q_0 = 0.028^{+0.36}_{-0.17}$), with the distribution showing no preference for an accelerating universe. For the SN-only analysis using the DES5Y sample, the estimated parameters, before the age-bias correction, are $w_0 = -0.82 \pm 0.13$, $w_a = -1.77^{+0.43}_{-1.21}$, and $q_0 = -0.26^{+0.12}_{-0.10}.$ After the age-bias correction, these values shift to $w_0 = -0.56^{+0.15}_{-0.12}$, $w_a = -1.75^{+0.47}_{-1.22}$, and $q_0 = 0.028 \pm 0.10.$ A similar trend is observed in the Pantheon+ dataset (see Table~\ref{tab:cosmo_params_2}). Before the age-bias correction, the estimated parameters for the combination of BAO and the Pantheon+ SN dataset are 
\[
\left.
\begin{array}{l}
w_0 = -0.89 \pm 0.06 \\[5pt]
w_a = -0.20 \pm 0.44 \\[5pt]
q_0 = -0.43 \pm 0.08
\end{array}
\right\} \quad \text{BAO+Panth.,}
\] 
and for the combination with DES5Y, they become
\[
\left.
\begin{array}{l}
w_0 = -0.78 \pm 0.07 \\[5pt]
w_a = -0.71 \pm 0.47 \\[5pt]
q_0 = -0.30 \pm 0.09
\end{array}
\right\} \quad \text{BAO+DES5Y.}
\] 
These values are almost identical to those reported in \citet{2025arXiv250314738D}. When combined with BAO measurements, the SN data before the age-bias correction reduce the apparent deviation from $\Lambda$CDM cosmology observed in BAO-only analyses. The inclusion of SN data without age-bias correction shifts the best-fit values of the dark energy parameters closer to those predicted by the $\Lambda$CDM model. Furthermore, the resulting deceleration parameter $q_0$ becomes clearly negative, favoring an accelerating universe. 

After the age-bias correction, however, the estimated values are substantially changed. For the combination with the corrected Pantheon+, they become
\[
\left.
\begin{array}{l}
w_0 = -0.45 \pm 0.07 \\[5pt]
w_a = -1.63 \pm 0.38 \\[5pt]
q_0 = 0.064 \pm 0.070
\end{array}
\right\} \quad \text{BAO+Panth. (corrected),} 
\] 
and for the combination with the corrected DES5Y, they are changed~to
\[
\left.
\begin{array}{l}
w_0 = -0.32 \pm 0.07 \\[5pt]
w_a = -2.14 \pm 0.39 \\[5pt]
q_0 = 0.199 \pm 0.070
\end{array}
\right\} \quad \text{BAO+DES5Y (corrected).}
\] 
The addition of SN data, after applying the age-bias correction, supports the deviation from the $\Lambda$CDM model indicated by BAO-only measurements, yielding best-fit values that are well aligned with the BAO-only case but with reduced uncertainties. The tension with the $\Lambda$CDM model exceeds $6\sigma$ for the BAO+DES5Y combination. Table~\ref{tab:cosmo_params_2} summarizes the estimated cosmological parameters, allowing a detailed comparison among BAO-only, and SN-only and BAO+SNe combinations with and without age-bias correction. For the combined datasets of both Pantheon+ and DES5Y, the application of the age-bias correction leads to a significant increase in $w_0$, consistent with the trend observed in the $w$CDM model, along with a substantial decrease in $w_a$, suggesting a stronger preference for a time-varying dark energy component. The correction also results in a marked increase in $q_0$, indicating a shift toward a non-accelerating universe.

Correcting for the age-bias in SN data resulted in a notable discordance among cosmological probes in both the standard $\Lambda$CDM and flat-$w$CDM models (see \citealt{2022MNRAS.517.2697L} and Section~\ref{s42}
). However, in a time-varying dark energy model, a new concordance emerges between BAO and SNe after the age-bias correction. To quantify the concordance among datasets, we compute the Kullback–Leibler (KL) divergence \citep{1320776d-9e76-337e-a755-73010b6e4b64}. This metric is well suited for our analysis, since cosmological parameter posteriors are often strongly non-Gaussian and influenced by priors, making traditional comparisons based on means or standard deviations unreliable. The KL divergence, as presented in \citet{2013PDU.....2..166V} is given by
\[
D_{KL}(\mathcal{P} \| \mathcal{W}) = \int_x \log_2 \left( \frac{\mathcal{P}(x)}{\mathcal{W}(x)} \right) \mathcal{P}(x) \, dx,
\]
where $\mathcal{P}$ is the distribution of interest and $\mathcal{W}$ is the reference distribution. A larger KL divergence indicates a greater difference between $\mathcal{P}$ and $\mathcal{W}$, with $D_{\mathrm{KL}} = 0$ indicating identical $\mathcal{P}$ and $\mathcal{W}$. In the upper panels of Figure~\ref{f7}, we show the results of the KL divergence, calculated from the marginalized posterior distributions of the parameters, with the BAO-only data serving as a reference. The combination with the Pantheon+ SN data is shown in the left panel, and that with the DES5Y SN data in the right panel. For all relevant parameters, the BAO+SNe combination exhibits lower KL divergences after applying the age-bias correction, indicating improved consistency with the BAO-only constraints.

This trend not only persists but becomes more pronounced when CMB data are combined with BAO and SN probes. Figure~\ref{f8} illustrates the impact of the age-bias correction on cosmological inferences derived from the combined BAO+CMB+SN datasets. The parameter estimates are also summarized in Table~\ref{tab:cosmo_params_2}. The combined analysis is categorized by the SN sample used (Pantheon+ or DES5Y) and by whether the age-bias correction is applied. The BAO+CMB only contours (gray) serve as a reference, highlighting changes in the posterior distributions resulting from the application of age-bias correction. It is important to note that the BAO+CMB only combination already suggests best-fit parameters ($w_0 = -0.43 \pm 0.21$, $w_a = -1.70 \pm 0.58$, and $q_0 = 0.075 \pm 0.215$) that significantly deviate from $\Lambda$CDM at the $\sim 3.0\sigma$ level, which is almost identical to the result from DESI DR2 ($3.1\sigma$). Our analysis indicates that these values have not changed much since the SDSS BAO \citep{2021PhRvD.103h3533A} and DESI DR1 results, although the associated uncertainties have been reduced. However, when the SN dataset before the age-bias correction is added here, the resulting SN-added contours lie closer to $\Lambda$CDM for both Pantheon+ and DES5Y. The estimated parameters for the combination with the Pantheon+ are
\[
\left.
\begin{array}{l}
w_0 = -0.84 \pm 0.05 \\[5pt]
w_a = -0.61 \pm 0.20 \\[5pt]
q_0 = -0.37 \pm 0.06
\end{array}
\right\} \quad \text{BAO+CMB+Panth.,}
\]
and for the combination with DES5Y, they become
\[
\left.
\begin{array}{l}
w_0 = -0.75 \pm 0.06 \\[5pt]
w_a = -0.85 \pm 0.22 \\[5pt]
q_0 = -0.27 \pm 0.06
\end{array}
\right\} \quad \text{BAO+CMB+DES5Y}.
\]
These values are fully consistent with those in Table 5 of \citet{2025arXiv250314738D}. Although the SN-added contours deviate from $\Lambda$CDM by $3.1\sigma$ (Pantheon+) and $4.4\sigma$ (DES5Y), as measured by the Mahalanobis distance in the ($w_0, w_a$) parameter space, they do not lie near the peak of the posterior distribution predicted soley from the BAO+CMB. Instead, they are located near the edge of the distribution, indicating a discordance between the BAO+CMB and the SN data before the age-bias correction. The SN dataset pulls the mean values of the dark energy equation-of-state parameters closer to the $\Lambda$CDM model.

However, after correcting the SN data for progenitor age-bias, the SN-added results shift back toward the BAO+CMB-only contours, aligning more closely with them for both the Pantheon+ and DES5Y samples. When the corrected Pantheon+ SN sample is combined with BAO+CMB, the best-fit values of $ w_0 $, $ w_a $, and $ q_0 $ are
\[
\left.
\begin{array}{l}
w_0 = -0.45 \pm 0.06 \\[5pt]
w_a = -1.59 \pm 0.23 \\[5pt]
q_0 = 0.065 \pm 0.060
\end{array}
\right\} \quad \text{BAO+CMB+Panth. (corrected),}
\]
and when the corrected DES5Y SN sample is added, they become
\[
\left.
\begin{array}{l}
w_0 = -0.34 \pm 0.06 \\[5pt]
w_a = -1.90 \pm 0.25 \\[5pt]
q_0 = 0.178 \pm 0.061
\end{array}
\right\} \quad \text{BAO+CMB+DES5Y (corrected)}.
\]
The right panels of Figure~\ref{f8} show that the posterior distributions for $ w_0 $, $ w_a $, and $ q_0 $, all agree more closely with those from BAO+CMB alone after applying the age-bias correction. The KL divergence, computed using the BAO+CMB-only combination as the reference (see the bottom panels of Figure~\ref{f7}), also decreases consistently across all relevant parameters when the age-bias correction is applied, indicating improved agreement among the probes. Thus, a new cosmological concordance emerges when the SN distance scale is corrected for age-bias within the $ w_0w_a $CDM model. The discordance with the $ \Lambda $CDM model now reaches at 9.8$ \sigma $ (Pantheon+) and 11.7$ \sigma $ (DES5Y), as quantified by the Mahalanobis distance in the ($ w_0 $, $ w_a $) parameter space, strongly suggesting a time-varying dark energy equation of state.

The SN datasets used in our analysis, DES5Y and Pantheon+, incorporate the host galaxy stellar mass step correction in SN magnitude standardization. This is a magnitude offset between hosts above and below $\sim 10^{10} M_{\odot}$, commonly parameterized by $\gamma$. As discussed in Section~\ref{s2}, the mass step correction cannot account for the redshift-dependent bias arising from progenitor age, as the evolution of progenitor age with redshift differs significantly from that of host galaxy mass. Nevertheless, it is important to assess the potential for over-correction when both the age-bias and host mass corrections are applied to SN magnitudes. We therefore performed this test using the DES5Y dataset, which adopts $\gamma = 0.038$. For the Pantheon+, the $\gamma$ value is apparently negligible, with $\gamma = -0.003$. In the $w_0w_a\mathrm{CDM}$ model using only the SN dataset, removing the host mass step correction (i.e., $\gamma = 0$) shifts the best-fit parameters by $\Delta w_0 = -0.036$, $\Delta w_a = 0.016$, and $\Delta q_0 = -0.015$. When BAO and CMB data are included, the corresponding shifts are $\Delta w_0 = 0.002$, $\Delta w_a = 0.005$, and $\Delta q_0 = 0.003$. Overall, the impact of the host mass step correction on our results is negligible relative to the impact of the age-bias correction and the typical MCMC posterior uncertainties.

\vspace{+6.2em}

\section{Discussion} \label{s5}

\begin{figure*}
\includegraphics[angle=0,scale=0.42]{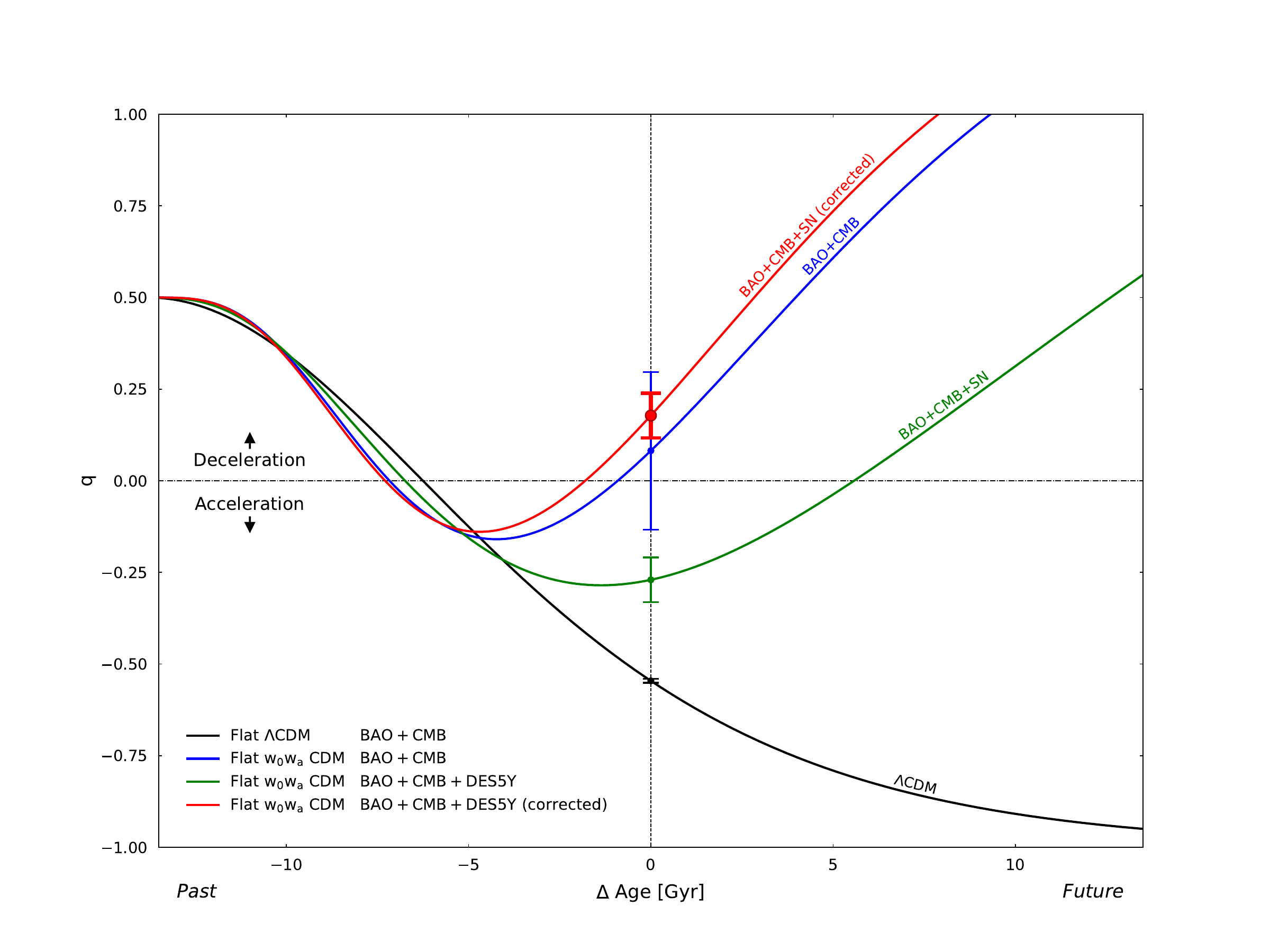}
\caption{
Evolution of the deceleration parameter, $ q = -\ddot{a}a/\dot{a}^2 $, of the universe. The vertical dotted line indicates the present age of the universe. The black line shows the result from the BAO + CMB data under the assumption of the flat-$ \Lambda $CDM model. The green and red lines represent the flat-$ w_0w_a $CDM models derived from the combined datasets (BAO + CMB + DES5Y SN) before and after applying the age-bias correction to the SN dataset, respectively. After the age-bias correction, the combined datasets suggest that the universe is already in a state of decelerated expansion ($ q_0 > 0 $) at the present epoch ($ \Delta{\rm Age}=0.0 \ \rm{Gyr}$) and will remain so in the future, consistent with the prediction from DESI BAO combined with CMB alone (the blue line).
\label{f9}}
\end{figure*}

\begin{figure*}
\includegraphics[angle=0,scale=0.40]{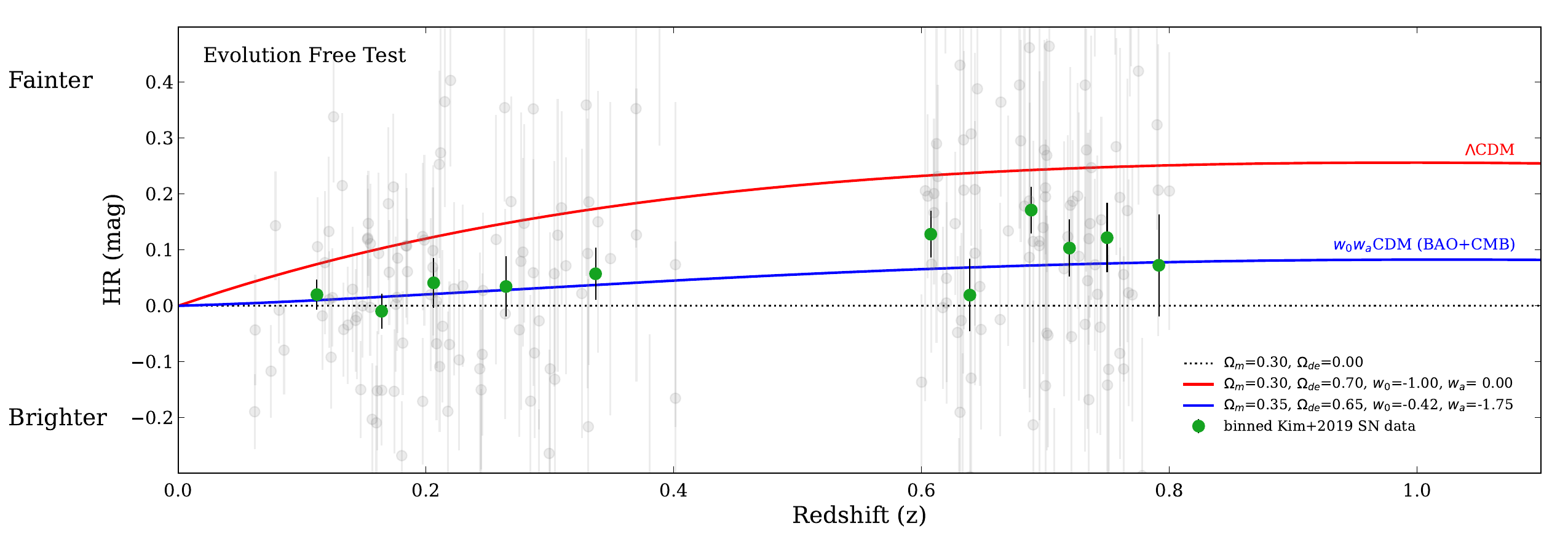}
\caption{
The evolution-free cosmological test in the residual Hubble diagram. Only SNe from young ($< 4.5$ Gyr) host galaxies are selected from the \citetalias{2019ApJ...874...32R} and \citetalias{2011ApJ...740...92G} samples at $0.05 < z < 0.4$, such that their mean stellar population age ($3.1$ Gyr) matches that of host galaxies at $z = 0.7 \pm 0.1$. The HR values are taken from \citet{2019JKAS...52..181K} with a 0.146 mag offset applied to account for the mean age difference between all host galaxies and the young hosts at $z \sim 0.0$ (gray open circles). The binned SN data (green circles) are more consistent with the $w_0w_a$CDM model favored by DESI BAO (DR2) and CMB combination (blue line) than with the $\Lambda$CDM model (red line).
\label{f10}}
\end{figure*}

Our analysis is fundamentally different from previous studies that incorporated progenitor age effects only indirectly through age proxies. For example, \citet{2020A&A...644A.176R} and \citet{2021A&A...649A..74N}, using the SN magnitude step with the local specific star formation rate, considered only the redshift evolution of the relative fractions of “young” and “old” progenitors, without accounting for the absolute age evolution of the “old” progenitors across redshift. As shown in Figure~\ref{f2}, the mean age of “old” progenitors declines markedly from ∼10~Gyr to ∼3~Gyr between $ z = 0.0 $ and $ z = 1.0 $. This systematic effect was not included in their analyses, leading to a significant underestimation of the full impact of progenitor age-bias. The DES SN program, as described by \citet{2024ApJ...975...86V}, employed the progenitor age model of \citet{2021MNRAS.506.3330W} solely to explore the correlation between the light-curve width parameter $ x_1 $ and progenitor age. However, this limited treatment was incorporated only into the systematic error budget, with its impact reportedly negligible (see their Table~7). The direct correlation between host age and HR, together with the redshift evolution of the mean progenitor age, as described in Sections~\ref{s2} and~\ref{s3}, has not been explicitly incorporated into SN cosmology analyses by other studies.

The most intriguing outcome of our study is that it offers a new perspective on the evolution of the deceleration parameter. Figure~\ref{f9} illustrates how the deceleration parameter evolves over time, before and after applying the progenitor age-bias correction to the SN data. For comparison, it also shows the evolution predicted by the $ \Lambda $CDM model, which has been the standard cosmological model until recently. As is well known, the $ \Lambda $CDM model predicts that the present universe is in a phase of accelerated expansion and will continue to accelerate in the future. In contrast, the evolution predicted when the DESI BAO result is combined with CMB and SN data shows that the future universe will transition to a state of decelerated expansion. Nevertheless, even in this case, the present universe remains in a phase of accelerated expansion ($ q_0 < 0 $). However, when the progenitor age-bias correction is applied to the SN data, not only does the future universe transition to a state of decelerated expansion, but the present universe also already shifts toward a state closer to deceleration rather than acceleration. Interestingly, this result is consistent with the prediction obtained when only the DESI BAO and CMB data are combined, as discussed in Section~\ref{s1}. Together with the DESI BAO result, which suggests that dark energy may no longer be a cosmological constant, our analysis raises the possibility that the present universe is no longer in a state of accelerated expansion. This provides a fundamentally new perspective that challenges the two central pillars of the $ \Lambda $CDM standard cosmological model proposed 27 years ago.

The progenitor age-bias in SN cosmology might also contribute to alleviating the Hubble tension \citep{2022ApJ...934L...7R}. This arises from the potential population mismatch between the host galaxies in the calibration sample and those in the Hubble flow sample. Currently, the calibrating galaxies in the second rung of the distance ladder are all late-type galaxies with relatively young stellar populations, whereas host galaxies in the Hubble flow sample (third rung) generally encompass all morphological types and, on average, have relatively older stellar populations \citep{2021ApJ...919...16F, 2011ApJ...730..119R}. More recently, \citet{2022ApJ...934L...7R} argue that they employ only spiral galaxies for the Hubble flow sample to minimize this potential population mismatch. However, simply selecting host galaxies based on morphological classification may not sufficiently ensure that the SNe~Ia in the third rung are identical to those in the second rung in terms of progenitor age. This is particularly relevant because \citet{2015ApJ...802...20R} demonstrated that most SNe~Ia in the second rung originate from locally star-forming environments, whereas, according to \citet{2016MNRAS.459.3130A}, only two-thirds of spiral galaxies host SNe~Ia on spiral arms (i.e., in locally star-forming, young environments). Therefore, if the conclusion of \citet{2015ApJ...802...20R} remains valid for the expanded sample of 37 late-type calibrating galaxies in \citet{2022ApJ...934L...7R}, we would still expect some population mismatch between the SNe~Ia in the second and third rungs. Furthermore, the accuracy of morphological classification is not complete and estimated to be $\sim90\%$ at $z \sim 0.1$ \citep{2005ApJ...635L..29P}, and, therefore, not all of the Hubble flow sample in \citet{2022ApJ...934L...7R} would be purely spiral hosts. Taken together, only $\sim60\%$ of SNe~Ia in the Hubble flow sample of \citet{2022ApJ...934L...7R} might originate from relatively young populations, whereas $\sim90\%$ of SNe~Ia in the calibration sample could be from such populations. The extreme case of the age-bias in the Hubble tension is expected between the calibration sample (mostly late-type galaxies) and the host galaxies in the Coma cluster, which comprises mostly early-type galaxies. Indeed, a recent study by \citet{2025ApJ...979L...9S} indicates that the Hubble tension is becoming even more severe in this case. If this possibility is confirmed by direct age measurements of stellar populations in host galaxies, even a 2--3~Gyr difference in the average age between the second and third rungs could significantly reduce (by 3 -- 4.5\%, $\Delta H_0 = 2.2$--$3.3$ km/s/Mpc) the value of the Hubble constant while increasing its systematic error. This, in turn, could substantially alleviate the Hubble tension.

The redshift-dependent age-bias correction applied in this study is in good agreement with the directly observed evolution of the mean stellar population age of galaxies (see Figure~\ref{f2}). Therefore, on average, this approach appropriately accounts for the systematic variation of the age-bias with redshift. To provide a more direct cosmological analysis free from the need for progenitor age-bias correction, we also perform an “evolution-free” cosmological test, where only SNe from young and coeval host galaxies are used across the full redshift range. This approach, suggested by several earlier studies \citep{1998AJ....116.1009R, 2016ApJS..223....7K}, yields cosmological constraints that are inherently free from the influence of the progenitor age-bias. Figure~\ref{f10} presents the first result from this test. To maximize the number of host galaxies within a consistent SN catalog, the HR values are taken from \citet{2019JKAS...52..181K}, in which the mass-step correction is not applied. Based on the CSFH and direct age measurements of galaxies (Figure~\ref{f2}), the mean stellar population age of SN host galaxies at $z = 0.7$ is estimated to be approximately 3.1~Gyr. We therefore select only comparably young ($< 4.5$~Gyr) host galaxies at relatively low redshifts from the \citetalias{2019ApJ...874...32R} and \citetalias{2011ApJ...740...92G} samples that overlap with \citet{2019JKAS...52..181K} catalog, such that their mean age also becomes 3.1~Gyr. As a result, the binned SN data (green circles) in Figure~\ref{f10} represent a sample with approximately coeval progenitor ages across redshift. It is important to confirm that these homogeneous data remain more consistent with the $w_0w_a$CDM model preferred by the combined DESI BAO and CMB analysis than with the $\Lambda$CDM model. Figure~\ref{f10} therefore provides independent support that strengthens the overall conclusion of this study. In future work, we plan to extend this evolution-free test by measuring stellar population ages for additional host galaxies in currently under-sampled redshift ranges.

\section*{Acknowledgments}

We acknowledge support from the National Research Foundation of Korea to the Center for Galaxy Evolution Research (RS-2022-NR070872, RS-2022-NR070525).

\section*{Data Availability}

This article makes use of BAO, CMB, and SN data which are publicly available at \url{https://github.com/CobayaSampler}. ACT DR6 data can be accessed at \url{https://github.com/ACTCollaboration/act_dr6_lenslike}. Hubble residuals of the YONSEI SN catalog of \citet{2019JKAS...52..181K} were obtained through private communication with Young-Lo Kim. The ages of the corresponding SN~Ia host galaxies are available at \url{https://academic.oup.com/mnras/article/538/4/3340/8098234}.





\bibliographystyle{mnras}
\bibliography{paper_II} 



\bsp	
\label{lastpage}
\end{document}